\documentclass[a4paper]{article}
\newskip\humongous \humongous=0pt plus 1000pt minus 1000pt

\newif\ifdtup

\catcode`\@=11

\@addtoreset{equation}{section}
\def\theequation{\arabic{equation}}

\def\@normalsize{\@setsize\normalsize{15pt}\xiipt\@xiipt
\abovedisplayskip 14pt plus3pt minus3pt%
\belowdisplayskip \abovedisplayskip
\abovedisplayshortskip \z@ plus3pt%
\belowdisplayshortskip 7pt plus3.5pt minus0pt}

\def\small{\@setsize\small{13.6pt}\xipt\@xipt
\abovedisplayskip 13pt plus3pt minus3pt%
\belowdisplayskip \abovedisplayskip
\abovedisplayshortskip \z@ plus3pt%
\belowdisplayshortskip 7pt plus3.5pt minus0pt
\def\@listi{\parsep 4.5pt plus 2pt minus 1pt
     \itemsep \parsep
     \topsep 9pt plus 3pt minus 3pt}}

\relax

\catcode`@=12

\catcode`\@=11
\def\section{\@startsection{section}{1}{\z@}{3.5ex plus 1ex minus
   .2ex}{2.3ex plus .2ex}{\large\bf}}
\def\thesection{\arabic{section}.}

\def\appendix{\setcounter{section}{0}
 \def\thesection{Appendix \Alph{section}:}
 \def\theequation{\Alph{section}.\arabic{equation}}}

\newcommand{\beq}{\begin{equation}}
\newcommand{\eeq}{\end{equation}}
\newcommand{\bea}{\begin{eqnarray}}
\newcommand{\eea}{\end{eqnarray}}

\newcommand{\non}{\nonumber}

\newcommand{\Tr}{\mbox{Tr}}

\def\br{{\tilde \beta}}

\def\r4{{\bf R}^4}
\def\s4{{\bf S}^4}

\def\Tr{ \hbox{\rm Tr}}
\def\const{\hbox {\rm const.}}
\def\o{\over}

\def\bra{\langle}
\def\ket{\rangle}
\def\Arg{\hbox {\rm Arg}}
\def\Re{\hbox {\rm Re}}
\def\Im{\hbox {\rm Im}}
\def\mathrm{\hbox{\rm}}
\begin{document}
\begin{titlepage}
\begin{center}
{\Large
   On   the  Beta Function  in \\
Supersymmetric Gauge Theories
}
\end{center}
\vspace{1em}
\begin{center}
{\large Giuseppe Carlino$^1$, Kenichi Konishi$^2$, Nicola Maggiore$^1$ \\
and \\Nicodemo Magnoli$^1$ }
\end{center}
\vspace{1em}
\begin{center}
{\it
Dipartimento di Fisica -- Universit\`a di Genova$^{(1)}$\\
Istituto Nazionale di Fisica Nucleare -- Sezione di Genova$^{(1)}$\\
Via Dodecaneso, 33 -- 16146 Genova (Italy)

Dipartimento di Fisica -- Universit\`a di Pisa$^{(2)}$\\
Istituto Nazionale di Fisica Nucleare -- Sezione di Pisa$^{(2)}$\\
Via Buonarroti, 2, Ed. B. -- 56127 Pisa (Italy)

E-mail: carlino@ge.infn.it; konishi@mailbox.difi.unipi.it; \\
maggiore@ge.infn.it; magnoli@ge.infn.it~~.
}
\end{center}
\vspace{3em}
\noindent
{\bf Abstract:}
{ We  re--examine perturbative and nonperturbative aspects of the beta function in $N=1$ and $N=2$ supersymmetric gauge theories, make comments on the recent literature on the subject and discuss the exactness of several known results such as the NSVZ beta function.}

\vspace{1.5em}
\begin{flushleft}
GEF-TH 2/99\\
IFUP-TH 10/99
\end{flushleft}
\begin{flushright}
February, 1999
\end{flushright}

\end{titlepage}

\newpage

\noindent {\bf 1.}
In this  note    several   comments  on the
beta function in supersymmetric gauge theories  will be made
 in the light of the  recent  literature on
the  subject \cite{AHM1,AHM2,AFY,n2betafns}.
The bare Lagrangian of an  $N=1$ supersymmetric gauge theory with generic
matter content is given by
\beq  L=  {1 \o 4} \int  d^2 \theta
\left( {1 \o g_h^2(M) }
\right) W^a W^a + \mbox{h.c.} +  \int d^4\theta \sum_i  \Phi_i^{\dagger}
e^{2V_i}  \Phi_i
\eeq
where
\beq   {1 \o g_h^2(M)}  = {1 \o g^2(M)} +  i\, {\theta(M)\o
8\pi^2}\equiv  {\tau(M) \o  4 \pi}
\label{holomorphy}
\eeq
and $g(M)$ and $\theta(M)$  stand for the bare   coupling constant and
vacuum parameter,   $M$ being the ultraviolet cutoff.
By  a   generalized   nonrenormalization theorem \cite{NSV}
the effective Lagrangian at scale $\mu$ has the form,
\beq
L=  {1 \o 4} \int  d^2 \theta
\left( {1 \o g_h^2(M) } + {b_0 \o 8 \pi^2} \log {M \o \mu}
\right) W^a W^a + \mbox{h.c.} +  \int d^4\theta \sum_i Z_i(\mu, M) \Phi_i^{\dagger}
e^{2V_i}  \Phi_i \, ,
\label{eflag1}
\eeq
(plus higher dimensional  terms). Here
\beq
b_0= 3 N_c- \sum_i T_{Fi}; \qquad T_{Fi}= {1 \o 2}  \qquad (\mbox{quarks})\, .
\eeq

Novikov et. al.  used then  the 1PI  effective action to define a
``physical'' coupling constant
for which they obtained the well known NVSZ beta function (Eq.(\ref{beta})
below) \cite{betafn}.

Recently the derivation  of the NVSZ beta function was  substantially
clarified by
Arkani-Hamed and Murayama \cite{AHM1,AHM2}.  (See also \cite{AFY}.)
 They work entirely in the framework of
the Wilsonian effective action (hence no subtleties due to zero momentum
external lines, such as
those leading to apparent violation of nonrenormalization
theorem\cite{NSV,Jones}).   They insist
simply that at each infrared cutoff $\mu$  the matter kinetic terms  be
re-normalized so that it resumes the standard canonical form.
Thus by introducing
\beq
\Phi_i = Z_i^{-1/2} \Phi_i^{(R)} \, ,
\eeq
and  by taking into account the appropriate anomalous Jacobian \cite{KS},
one gets
\bea
L &=&  {1 \o 4} \int  d^2 \theta
\left( {1 \o g_h^2(M) } + {b_0 \o 8 \pi^2} \log {M \o \mu}
 - \sum_i {T_F \o 8\pi^2} \log Z_i(\mu, M) \right) W^a W^a + \mbox{h.c.} \non\\
 &+& \int d^4\theta \sum_i  \Phi_i^{(R) \dagger} e^{2V_i}  \Phi_i^{(R)}
 \non \\ 
&=& {1 \o 4} \int  d^2 \theta
{1 \o g_h^2(\mu) } W^a W^a + h.c.+ \int
d^4\theta \sum_i  \Phi_i^{(R) \dagger} e^{2V_i}  \Phi_i^{(R)} \, .
\eea
where
\beq
{ 1\o g_h^2(\mu)} \equiv
{1 \o g_h^2(M) } + {b_0 \o 8 \pi^2} \log {M \o \mu}
 - \sum_i {T_{Fi} \o 8\pi^2} \log Z_i(\mu, M) \, .
\label{ghmu}
\eeq
This leads to the beta function (call it $\beta_h$ to distinguish
it from the more commonly used definition):
\beq
 {\beta_h}(g_h) \equiv  \mu {d \o d \mu}  \Re \, g_h = - { g_h^3 \o 16
\pi^2} \left(
3 N_c -  \sum_i T_{Fi} (1- \gamma_i) \right) \, ,
\label{tildebeta}
\eeq
where
\beq
\gamma_i(g_h(\mu) ) =   - \mu { d \o d \mu } \log Z_i(\mu, M)
\eeq
is the anomalous dimension of the $i-$th matter  field.
The same result follows by differentiating (\ref{ghmu}) with respect to $M$
with $\mu$ fixed, and by using
 $\gamma(g_h(M) ) =   + M { d \o d M } \- \log Z_i(\mu, M)$.

The ``holomorphic'' coupling constant   $g_h(\mu) $    is a perfectly good
definition
of the effective coupling constant:  it is finite as
$   M \to \infty; \quad \mu = {\hbox{\rm finite}}, $
and  physics  below $\mu$ can be computed in terms of it.
On the contrary,   the coefficient of $W^a W^a$  in (\ref{eflag1})  is {\it
not} a good
definition of an effective coupling constant,  as long as $N_f \ne 0$:  it
is divergent in the
 limit the ultraviolet cutoff is taken to infinity.
Let us note that,  in spite of its  name, the holomorphic coupling constant
gets renormalized in a non-holomorphic way, due to the fact that $Z_i(\mu, M)$
is real.

Finally,  in order to have  the canonical form of gauge kinetic terms,
$ -F_{\mu \nu}F^{\mu \nu}/4, $  one must  perform a further   change of the
variables,
\beq
A_{\mu}= g_c A_{c \mu} \, ,
\eeq
and  the  corresponding
rescaling of   the gaugino field   $\lambda_{\alpha}
(x)$,  to preserve  supersymmetry.  This introduces as functional--integral
Jacobian   an extra factor \cite{AHM2},
\beq
\exp  \, {1 \o 4} \int d^4x  \int d^2\theta   \,  {N_c  \log g_c^2
\o 8 \pi^2 } \,  \, W^a W^a + \mbox{h.c.}
\eeq
and as a consequence,   leads to  the change of the coupling constant
\cite{SV91},
\beq
\Re {1 \o g_h^2} = {1 \o g_c^2} + { N_c  \o 8 \pi^2 } \log g_c^2 \, .
\label{gcanon}
\eeq
The NSVZ beta function \cite{betafn} follows then from (\ref{gcanon}) and
(\ref{tildebeta}):
\beq
\beta(g_c)=- { g_c^3 \o 16 \pi^2}
{ 3 N_c -  \sum_i T_{Fi} (1- \gamma_i) \o
1- N_c g_c^2/8\pi^2} \, .
\label{beta}
\eeq
 An important point of   \cite{AHM2}    is the fact that
the Wilsonian  coupling constant,  whether  ``holomorphic'' or  ``canonical'',
contains  higher loop  perturbative corrections in general;   the often
stated one-loop
(perturbative)  exactness of the Wilsonian effective coupling constant is
valid only in particular cases,
e.g., for the ``holomorphic''  coupling constant in the  $N=1$
 supersymmetric  pure   Yang-Mills theory.

\bigskip

\noindent {\bf 2. }
The above procedure    nicely  ``explains'' the origin of the
denominator of the NSVZ beta
function.    In the case of $N=1$
   pure   Yang-Mills theory
the latter    has led to an interesting  conjecture \cite{denombeta}.
  However,
it also leads to  a new puzzle.    In fact, the right hand side of
(\ref{gcanon}) has a minimum at
$g_c^2= 8\pi^2/N_c$,  precisely corresponding to  the pole of the NSVZ beta
function,   where it takes the
value,
\beq
{N_c \o 8 \pi^2 } \log{8 \pi^2 e \o N_c} \, ,
\eeq
which is positive  unless $N_c$ is rather large (i.e., unless $N_c \ge 215$).
On the contrary the left hand side of
(\ref{gcanon}) evolves down to zero if the beta function has no zero
($N_f < 3 N_c /2$).   Thus
for large values of $g_h$ ($g_h  >    8\pi^2/N_c \log(8 \pi^2 e/N_c)$)
the redefinition
(\ref{gcanon}),   with a real ``canonical coupling constant'',   is not
allowed.

Since this problem occurs  for any $N_f$  such that  $N_f < 3 N_c /2$,
let us  for simplicity
consider the case of the
pure  Yang Mills  theory ($N_f=0$) and compare the RG evolution in the two
coupling constants.  In this case,
${\beta_h}(g_h) $ is  a pure one-loop effect,  so that RG equation can be
integrated in a closed form:
\beq
{1 \o 2 g_h^2(\mu)} - {3 N_c \o 16 \pi^2} \log \mu =
\{ \mbox{indep. of } \mu\} \,\,\equiv  - {3 N_c \o 16 \pi^2} \log
\Lambda \, ,
\eeq
(which defines $\Lambda$)  namely,
\beq
{ 1\o g_h(\mu)^2} =  {3N_c \o 8 \pi^2} \log {\mu \o \Lambda} \, ,
\label{deflambda}
\eeq
which evolves to the infrared and  vanishes at  $ \mu= \Lambda$.

On the other hand, if one integrates
\beq
\mu { d \o d \mu} g_c=  \beta(g_c)=- { g_c^3 \o 16 \pi^2}
{ 3 N_c  \o
1- N_c g_c^2/8\pi^2} \, ,
\eeq
one gets
\beq
{1 \o  g_c^2(\mu)} +  {N_c \o 8 \pi^2} \log g_c(\mu)^2  - { 3N_c \o 8
\pi^2 } \log \mu = \{\mbox{indep. of  } \mu \} \, .
\eeq
so
\beq
{1 \o  g_c^2(\mu)} +  {N_c \o 8 \pi^2} \log g_c(\mu)^2 =
 { 1\o g_h(\mu)^2} =  {3N_c \o 8 \pi^2} \log {\mu \o \Lambda} \, ,
\label{rgforbcan}
\eeq
by using the same $\Lambda$ as in Eq.(\ref{deflambda}).
The problem with (\ref{rgforbcan}) is that   $g_c(\mu)$  does not
``run'' down to $\mu = \Lambda$;   it runs only  down  to
\beq
\mu_0=  {8 \pi^2 e \Lambda \o N_c}   > \Lambda \, ,
\eeq
(for $8 \pi^2 e >  N_c$),     which corresponds to the pole of the NVSZ
beta function.

Also, (\ref{rgforbcan})  apparently   suggests  the presence of another
branch in which the coupling
constant $g_c$ grows in the ultraviolet \cite{denombeta}.

Actually   both the absence of evolution below the scale $ {8 \pi^2 e
\Lambda / N_c}
$ and  the apparent new phase of the theory  are  probably artefacts
 caused by   the illegitimate change of
variable (\ref{gcanon}).
The pole at $g_c^2=8\pi^2/N_c$ is  then simply  a sign of the failure of
$g_c$ as a coupling constant (and $A_{c \, \mu }(x)$  as a functional
variable),
a sort of a singularity of parametrization,  rather than of
 physics itself.

This however means that if one starts at high energies  by using the
standard ``canonical'' coupling constant
and  studies the  RG evolution towards the low energies,   one must switch
to the ``holomorphic''
description  at certain point (in any case,   before the ``critical'' value
$g_c^2= 8\pi^2/N_c$ is reached),
   in
order to describe the physics smoothly down to $\mu = \Lambda.$
 The impossibility of writing  a low energy effective Lagrangian with
canonically normalized
gauge kinetic terms,  does not represent   any inconsistency
  since
the  low energy  physical degrees of
 freedom   are in fact  described by    some gauge-invariant composite
fields,   and   not by  gauge fields themselves.  The latter
 fluctuate violently while the  appropriate variables behave more smoothly.
 For  $N_f \le N_c$,   the appropriate  low energy degrees of freedom  are
mesonlike
composite fields  $M_{ij} ={\tilde Q}_i  Q_j  $;\footnote{As is well known
SQCD with   $N_f   <   N_c$   does not have well defined vacua  if quarks
are massless:
we assume that all quark masses are small but nonvanishing.  In the case
 $N_f   =  N_c=N, $ the low energy degrees of freedom contains the baryon
 $B= \epsilon_{a_1 a_2 \ldots a_{N} }\epsilon^{
i_1  i_2 \ldots  i_{N} }   Q_{i_1}^{a_1}\,  Q_{i_2}^{a_2} \dots
Q_{i_{N}}^{a_{N}} $
and  ${\tilde B}$ defined analogously  in terms of ${\tilde Q}$' s,      as
well.)
}    for $N_f=N_c+1$  they are the
mesons and  ``baryons'';  for $N_c +1 \le N_f \le 3 N_c$
 the low energy degrees
of freedom are   dual quarks  and free mesons \cite{Seiberg1}.  In
particular,  in the conformal
window,  i.e.~for   $3 N_c /2 \le N_f \le 3 N_c$,  the  low energy theory
admits two  dual equivalent
descriptions,  one in terms of the original quarks and gluons of $SU(N_c)$
gauge theory,
 another in
terms of  $SU(N_f- N_c)$  gauge theory with $N_f $ flavors of dual quarks.

At larger values of $N_f$  ($N_f > 3 N_c$)  the low energy degrees of
freedom are
the original quarks and gluons,  but  since the theory is infrared free no
obstruction arises  against
describing them by using  the canonical coupling constant at all scales.

 The success of the NSVZ beta function in the case  of SQCD
in the conformal window ($3 N_c/2  < N_f < 3 N_c$) \cite{Seiberg1},
especially the
determination of the anomalous dimension of the matter fields,
\beq
\gamma^{*}= { 3 N_c- N_f \o N_f} \, ,
\label{andim}
\eeq
at the infrared fixed point,
 is consistent with  the use of the holomorphic coupling constant
(Eq.(\ref{tildebeta})).    It does neither  require the use of the canonical
coupling constant nor  necessitate  the form of the original
NSVZ beta function.    This is important because
the  anomalous dimension at the IR fixed point is a physically observable
number.

\bigskip

\noindent  {\bf  3. }
One might wonder how ``exact'' all this  is.  It is clear that the
diagrammatic  proof of
the  generalized  nonrenormalization theorem of \cite{NSV}  is  valid only
within perturbation
theory.

It was argued  on the other hand in \cite {AHM2}  that due to the existence
of an anomalous $U_R(1) $ symmetry
 the   beta  functions are purely
perturbative,  hence   the NSVZ beta function is exact perturbatively and
nonperturbatively,
at least for pure $N=1$  Yang--Mills theory.
 In fact,   the (holomorphic)   coupling constant at scale $M^{'}$  must
satisfy
\beq
{8 \pi^2\o g_h^2(M^{'})} =    {8 \pi^2\o g_h^2(M)} +   f\left({8
\pi^2\o g_h^2(M)}, t\right) \, ,
\eeq
where $t\equiv  \log {M\o M^{'}}$ and
$f$ is a holomorphic function of $g_h$.
It follows that
\beq
{\beta_h}(g_h) = (d / d \log M) \,
g_h(M)|_{g_h(M^{'})}
\eeq
shares the same property.    Together with the
periodicity in $\theta$
with period $2 \pi$, one finds that
\beq   {d \o dt}  { 8 \pi^2\o g_h^2(M)} =  - {16 \pi^2 \o  g_h^3}
{\beta_h}(g_h)
=   \sum_{n=0}^{\infty}  a_n  e^{- 8 \pi^2 n /g_h^2},
\eeq
where $a_n$ is the $n$- instanton contribution.   Since   the right hand
side  is independent of $\theta$  it must consists only of  the
perturbative term, $n=0$.

This   argument is however only valid  in theories
in which   $CP$
  invariance is not spontaneously broken.
Examples are  the pure $N=1$ Yang--Mills theory or  the    $N=1$ SQCD at the
origin of the space of
vacua   (with all scalar vevs vanishing):   there   the argument of
\cite{AHM2}   is valid and
the use of the NSVZ  beta function is justified.

Actually,  the argument can be reversed and used in a stronger manner.
In a generic   point of the space of vacua
 of $N=1$     SQCD,  or    of  a pure $N=2$   supersymmetric
Yang--Mills theory (a $N=1$ supersymmetric gauge  theory with a matter
chiral multiplet
in the adjoint representation),   for example,
 $CP$ invariance is spontaneously broken \cite{Konishi}, and there is a
nontrivial $\theta$ dependence.    By holomorphic dependence of $\beta$ on
$\tau= {\theta \o
2\pi } + {4\pi i\o  g^2}$  this implies that the beta function gets
necessarily   instanton corrections.
A na\"{\i}ve   application of the  NSVZ  formula  to the $N=2$ pure
Yang--Mills  theory,  which  would simply
yield      the purely one-loop perturbative   beta function
``$\beta(g)$''$=  -g^3/4\pi^2$,
is thus  incorrect.\footnote{Note that the original derivation of the  NSVZ
beta function
based on the calculation of certain  one--instanton amplitude,   applied in
a simple--minded way  to
this theory,   would yield    the one-loop beta function.   But this
argument  also  fails  since in the
presence of the adjoint scalar vev, the standard instanton selection rules
do not apply. }

\bigskip

\noindent {\bf 4. }   It might be thought  that
in the $N=2$
gauge theories  where    the exact low--energy effective action is known
\cite{SW,others},
the   exact  (nonperturbative) beta function can be computed.    For such
an attempts
see \cite{n2betafns}.   For concreteness   we  restrict our discussion
below   to the simplest such case:
the pure    $N=2\, $     $SU(2)$   Yang--Mills theory.

Due to the holomorphic
nature of Wilsonian effective action  the RG equation can be cast
into the form
\beq
\beta(\tau)\equiv   \mu { d \tau\o d \mu}  = { 2 i \o \pi} \,( 1 +
 c_1 \, e^{2 \pi i  \tau} + c_2 \, e^{4 \pi i  \tau}+ \ldots)
\label{RG1}
\eeq
for $\,  \Im \tau \gg 1\,$   (or $\,g^2 \ll 1\,$)  where
\beq
\tau= { \theta \o 2 \pi} + { 4 \pi i \o g^2} \, ,
\eeq
and  $\mu$ is the scale.
Written separately for the real and imaginary parts, Eq.(\ref{RG1}) reads
\cite{KnMo,Seiberg2}:
\bea
\beta_g(g, \theta) &\equiv&   \mu { d g\o d \mu}  =-{g^3\o 4 \pi^2} (
1+ c_1 \cos \theta e^{-8 \pi/g^2 } + \ldots) \, ; \non \\
\beta_{\theta}(g, \theta) &\equiv&   \mu { d \theta\o d \mu}
= -4 c_1 \sin \theta e^{-8 \pi/g^2 } + \ldots \, .
\eea
The coupling constant and the $\theta$ parameter will evolve in the infrared
up to the scale $\mu_{IR}$
which can be identified   with the mass of the lightest charged particle.
The difficulty   in finding the beta function in these theories    lies in
the fact that in general
the relation between  $\mu_{IR}$  and the gauge invariant  vev
$   u=  \bra Tr \phi^2  \ket $
is not simple.   (For particular cases see below.)  For this reason  the
knowledge of
\beq
\tau_{eff}(u)   =   {d a_D  \o da} \, ,
\eeq
from the exact Seiberg-Witten solution  as a function of the vacuum
parameter $u$, is not sufficient
to deduce  the correct $\beta$ function.

By integrating Eq.~(\ref{RG1}) one  gets:
\beq
\int^{\tau}  {d \tau \o \beta(\tau)} - \log \mu = C \, ;
\label{sol1}
\eeq
where $C$ is a $\mu$-independent integration constant.
The lower limit of the integration is
left unspecified: to change it is equivalent to a shift of $C$ by a
constant.
We set now $\mu= M$ ($M$ is the UV cutoff) and use the  known asymptotic
behaviour of $\tau$ and $\beta(\tau)$.
Since   the theory at large $u$ is weakly coupled at all scales,
one can use  the known behaviour of $\tau_{eff}$ for such cases \cite{ferrari},
\beq
\tau_{eff} \simeq   {i \o \pi} \log { 4 a^2 \o \Lambda^2} \, ,
\qquad   \theta_0  \equiv  -4   \Arg \, a
\eeq
where $\Lambda$ (real)  is defined such that  the  massless monopole occurs
at $u=\pm \Lambda^2$,
to find the behavior of $\tau(M)$ as a function of
$M$.   In fact, by identifying $\mu_{IR}=M= \sqrt 2  \, |a|$, one gets
\beq
\tau(M) \simeq   {\theta(M) \o 2\pi} +   {i \o \pi} \log { 2 M^2  \o
\Lambda^2}; \qquad
\beta(\tau) \simeq     {2 i \o \pi}
\label{tauasym}
\eeq
so  that
\bea
C &=& -{i \pi \o 2} \tau(M) - \log M + \const   \non \\
&=& -{i \pi \o 2}({\theta_0 \o 2 \pi} + {i \o \pi} \log{ 2 M^2 \o \Lambda^2} )
 - \log M  + \const    =  -\log { e^{i\theta_0/4}\o  \Lambda } + \const \, .
\label{Csol}
\eea
This shows how  the dynamical mass scale $ \Lambda $   and the bare
$\theta$ ($=\lim_{M \to \infty}
\theta(M)$ )
enter  together  as two integration constants of the RG equation.

In the strong coupling region, i.e.~near $\tau
= 0$ (which is explored by theories near   $u =\Lambda^2 $)  the infrared
scale is given by   $\mu_{IR}
=\sqrt 2 \, |a_D|$ (the monopole mass).   The
divergent  behavior of  the second term of the left hand side of
Eq.~(\ref{sol1})   must  be
cancelled by the first term:
\beq
\int^{\tau} d  \tau {1 \o \beta(\tau)} \simeq   \log {a_D \o
\Lambda} \, .
\eeq
>From the behavior of $a_D$ near $u=\Lambda^2$ \cite{SW}:
\beq
{a_D \o \Lambda} \simeq  i { u - \Lambda^2 \o 2 \Lambda^2}  =  16 i   q_D=
16 i  e^{i \pi \tau_D} \, ,
\eeq
one finds
\beq
\int^{\tau}d \tau {1 \o \beta(\tau)} \simeq   i \pi \tau_D \, ,
\label{reqrg}
\eeq
or  (by using $\tau = - 1/\tau_D$):
\beq
\beta(\tau)   \sim {1 \o  i \pi \tau_D^2}  =  -   {i \o  \pi}  \tau^2
\quad   \quad \rm{ as }
\quad \tau \to 0 \, .
\label{truebeta}
\eeq
For  CP invariant cases ($\theta=0$)  this means the behavior
\beq  \beta(g)\sim    -  {2 \o g},\eeq
at large $g$.

The existence of the nontrivial   space of vacua  implies that,
at any given scale,    besides the usual parameters
$g(\mu)$, $\theta(\mu)$   one has   the scale dependent vev,
\beq
v(\mu)=\bra \Tr \phi^2(\mu) \ket \,
\eeq
as another parameter of the theory.
The usual moduli parameter $u$ is to be identified with its value in the
low energy limit.
Note that even though $v(\mu)$ is complex,   its phase  is related
by  anomaly  to  $\theta(\mu)$ so that only three parameters
are independent.
At the UV cutoff this relation is the standard one:
\beq
v \to e^{i \alpha} v \quad \Longleftrightarrow  \quad  \theta \to
\theta + 2\alpha \, ,
\eeq
so that only the combination $\theta_{phys}=\theta -2 \, \Arg \,  v$ has a
physical
meaning.

The scale dependence of $v$ arises  because in the instanton
contributions to it the  integrations over the collective coordinates must
be done so that only
the distances between $1/M$ ($M$ being the UV cutoff) and $1/\mu$  are
involved.  For instance,
in the one instanton contribution the integration over the instanton size
must be limited to the
region,
\beq
{ 1 \o M} \le   \rho \le   { 1\o  \mu} \, .
\eeq
One is   thus led to  write  one more RG equation besides Eq.~(\ref{RG1}):
\beq
 \mu { d v \o d  \mu} = 2 v G(\tau) \, .
\label{RG2}
\eeq
When  the moduli parameter $u=\bra \Tr \phi^2 \ket$ is large as compared
to $\Lambda^2$, the theory is weakly coupled at all scales, and
$\sqrt2  |a| $ can be taken as the lower cutoff   $\mu$.   At large  $\mu$
(at $\tau \to i \infty$)   then
\beq
G(\tau) = 1  -   16\,  e^{2 i \pi \tau} + \ldots,  \qquad   G   < 1
\label{Gasym}
\eeq
from the  known  instanton expansion
\beq   u(a) = a^2   \sum_{n=0}^{\infty}    b_n   \left({\Lambda  \o
a}\right)^{4n},
\qquad \tau   \simeq   {i \o \pi} \log { 4 a^2 \o \Lambda^2}  \eeq
where  $b_0=1/2, \,\,b_1=1/4,$ etc.

Integrating Eq.~(\ref{RG2}) as before  by using  $d \mu /\mu= d \tau/
\beta(\tau)$,   one gets:
\beq
2 \int^{\tau} { d \tau \o \beta(\tau)}  G(\tau) - \log  {8 \bra \Tr
\,\phi^2 (\mu)
\ket \o \Lambda^2}   = R \, ,
\label{sol2}
\eeq
where $R$ is another  $\mu$--independent integration constant (the factor
$8/\Lambda^2$ has been
inserted for convenience).   Again, by taking $\mu=M$ (large),  using
Eq.~(\ref{tauasym}) and
Eq.~(\ref{Gasym}) as $\tau \rightarrow i \infty$, one gets \beq
R = -{i \pi }({\theta_0\o 2 \pi} + {i \o \pi} \log {2 M^2 \o \Lambda^2} )-
\log {8 \bra \Tr \, \phi^2 (M) \ket
\o \Lambda^2}= - \log {4  \, |\bra \Tr \phi^2 (M) \ket| \o M^2} \, .
\label{Rsol}
\eeq
Note that $R$ is real:  one finds thus the third    integration constant ${
|\bra \Tr \phi^2 (M) \ket| \o
M^2}$   besides $\theta_0$ and $\Lambda$.  They are the free
parameters of the theory.

By differentiating the left hand side of Eq.(\ref{sol2})  with respect to
$\tau$,  one gets
\beq
2 {G(\tau) \o \beta(\tau) } -  { d v \o d \tau} {1 \o v}=0 \, , \qquad
{\hbox {\rm or}}\qquad
2 v { d \tau \o d v}  = { \beta(\tau) \o G(\tau)} \, .
\label{veqn}
\eeq
Going to the IR limit, i.e.~$v \rightarrow u$, Eq.~(\ref{veqn}) reads
\beq
2 u { d \tau \o d u}  = { \beta(\tau) \o G(\tau) } \, ,
\eeq
where $\tau=\tau_{eff}$.
But the left hand side, which is the derivative of the low-energy effective
$\tau$
with respect to the vacuum parameter  $u$, can be computed from the knowledge
of the low energy actions only:  the result is  \cite{n2betafns}
\beq
2 u { d \tau \o d u} = { \beta(\tau) \o G(\tau)}  = {i \o \pi} ( {1 \o
{\theta_3}^4}
+ {1 \o {\theta_4}^4} )  \equiv \br(\tau)\,
\label{betaritz}
\eeq
where $\theta_i(\tau) = {\cal  \theta}_i(0 | \tau)$,  in terms of the
standard  elliptic
 theta functions \cite{ww}.

 $\br(\tau)$  does not represent  the nonperturbative beta function,  in
spite of
the  claim made in the literature to that effect, as can be seen from its
behavior near $\tau=0$, for instance.
>From the known properties of the theta functions
\bea
 { 1\o  \br(\tau) }   & = &   -i \pi   {    {\theta_3}^4(-1/\tau_D) \,
{\theta_4}^4(-1/\tau_D) \o  {\theta_3}^4(-1/\tau_D) + {\theta_4}^4(-1/\tau_D) }
 =  -i \pi (-i \tau_D)^2   {  {\theta_3}^4(\tau_D)  \, {\theta_4}^4(\tau_D) \o
{\theta_3}^4(\tau_D) +    {\theta_4}^4(\tau_D)   }   \non \\
&\simeq &   16 \pi   i \, \tau_D^2 \, e^{2 \pi i \tau_D}, \qquad
{\hbox{\rm as }} \quad  \tau_D \to i \infty \, ,
\eea
hence
\beq   \br(\tau) \simeq   -{ i \o 16 \pi}  \tau^2   e^{-2 \pi i/\tau},
\label{behtau0}\eeq
which differs from the  correct behavior of the beta function at $\tau \to 0+ i
\epsilon,$ Eq.~(\ref{truebeta}).

On the other hand,  Eq.~(\ref{behtau0}) is perfectly consistent with the
behaviour of $\tau$ as  a function of $u$.
\beq
2  \int^{\tau} { d \tau \o \beta(\tau)}  G(\tau)= 2
 \int^{\tau} { d \tau \o \br(\tau)} = 2 \int^{\tau_D} {d \tau_D \o
{\tau_D}^2 } {1 \o \br(\tau) }
\simeq  32 q_D = 32 e^{i \pi \tau_D} \, .
\label{ritrg}
\eeq
>From the expansion  of $a(u)$ as $u \to \Lambda^{2}$ \cite{ferrari}, one
finds:
\beq
\tau_D = - {da \o da_D} \simeq    - {i \o \pi} \log {u- \Lambda^2 \o 32
\Lambda^2} \, .
\eeq
Consequently:
\beq
2  \int^{\tau} { d \tau \o \beta(\tau)}  G(\tau)= \const +  {u-
\Lambda^2 \o \Lambda^2} \, .
\eeq
On the other hand,   the second term of Eq.~(\ref{sol2}) gives
\beq
- \log {8 u \o \Lambda^2} = \const - {u- \Lambda^2 \o \Lambda^2} \, ,
\eeq
near $u = \Lambda^2 $,
so that the linear dependence on $u- \Lambda^2$ cancels out, as it should.

One  can also check the behavior near  $\tau= \tau_0 = (1+i)/2$ of
Eq.~(\ref{sol2}),
which corresponds to the infrared behavior of  the theory near $u=0$.
By a simple calculation   using  the results of Ritz \cite{n2betafns},
one finds that:
 \beq
{ \beta(\tau) \o G(\tau)}=\br(\tau) \simeq   2 (\tau - \tau_0) \, .
\eeq
Thus
\beq
2 \int^{\tau} {d \tau \o \beta(\tau)}  G(\tau)  \simeq
\log (\tau - \tau_0) \, ,
\eeq
 which is singular.
But this singularity is expected because  the second term of Eq.~(\ref{sol2})
is also logarithmically divergent, since:
\beq
u \simeq    \pi i   {\theta_3}^4(\tau)|_{\tau_0} \cdot (\tau - \tau_0) \, ,
\eeq
as can be shown   by using the relation between  $u$ and   $\tau$:
\beq
u = { {\theta_3}^4(\tau) + {\theta_4}^4(\tau) \o {\theta_3}^4(\tau) -
{\theta_4}^4(\tau)} \, .
\eeq
Therefore, the logarithmic singularities cancel out.

These discussions simply check the derivation of (\ref{betaritz}),
but also shows   for instance  that
 the zero of
$\br(\tau) = {\beta(\tau) / G(\tau)}$  at    $ \tau_{0} = (1 + i)/2$ must
be attributed
  to a pole in the renormalization factor   $G(\tau)$,  not to a zero of the
beta function.    For if the beta function had a zero at  $ \tau_{0} = (1 +
i)/2$
the first term of the right hand side of  Eq.(\ref{sol1}) would  be
singular there:  such a singularity
would have to
be cancelled  by the second term,  which is however regular there because
the infrared cutoff
of the theory with $u=0$   is finite.  This is another way of saying that
the point $u=0$ is
not a special point in the space of vacua:  no restoration of the $SU(2)$
gauge symmetry occurs.
The fact that  $u\equiv  \bra \Tr \,
\phi^2\ket =  0   $  there,  is due to the
  instanton--induced renormalization of the composite operator $\Tr \,
\phi^2$.

In conclusion, the problem of finding the correct nonperturbative beta
function in supersymmetric gauge
theories   remains open.  Let us also note that the related issue of the
direct check of the
Seiberg--Witten formulas in various $N=2$  gauge theories by direct
instanton calculations,  after the
initial impressive success,
leaves still many  questions unanswered \cite{DKM}.

\bigskip
\noindent {\bf Ackowledgment}
One of the authors (K.K.)  thanks M. Sakamoto, H. Murayama and F. Fucito
for useful discussions at various  stages of the work.

\end{document}